\documentclass[
  journal=pasa,
  manuscript=research-paper, 
  year=202X,
  volume=YY,
]{cup-journal}

\usepackage{xcolor} 
\usepackage{hyperref}
\hypersetup{
    colorlinks=true,
    linkcolor=blue,
    citecolor=violet,
    urlcolor=violet
}
\usepackage{smartdiagram}
\usesmartdiagramlibrary{additions}
\usepackage{longtable}
\usepackage{amsmath}
\usepackage{cprotect}
\usepackage{float}
\usepackage{graphicx}
\usepackage{overpic}
\usepackage{tabularx}
\usepackage{makecell}
\usepackage{lipsum}

\usepackage{subcaption}
\usepackage{amssymb,microtype,siunitx,booktabs}
\newcommand{\bdfull}{WISE\allowbreak~J151721.13\allowbreak$+$052929.3}
\newcommand{\bdshort}{WISE\allowbreak~J1517\allowbreak$+$05}
 
\newcommand{\askapfull}{ASKAP\allowbreak~J151721.1\allowbreak$+$052931}
\newcommand{\askapshort}{ASKAP\allowbreak~J1517\allowbreak$+$0529}

\title{Periodic radio emission from a rapidly rotating T8 ultracool dwarf}

\author{Iris de Ruiter}
\affiliation{Sydney Institute for Astronomy, School of Physics, The University of Sydney, NSW 2006, Australia}
\alsoaffiliation{ARC Centre of Excellence for Gravitational Wave Discovery (OzGrav), Australia}
\email[Iris de Ruiter]{\url{iris.deruiter@sydney.edu.au}}

\author{Joshua Pritchard}
\affiliation{Australia Telescope National Facility, CSIRO Space and Astronomy, PO Box 76, Epping, NSW 1710, Australia}

\author{Tara~Murphy}
\affiliation{Sydney Institute for Astronomy, School of Physics, The University of Sydney, NSW 2006, Australia}
\alsoaffiliation{ARC Centre of Excellence for Gravitational Wave Discovery (OzGrav), Australia}

\author{Andrew K. Hughes}
\affiliation{Department of Physics, University of Oxford, Denys Wilkinson Building, Keble Road, Oxford OX1 3RH, UK}

\author{David L. Kaplan}
\affiliation{Department of Physics \& Astronomy, University of Wisconsin-Milwaukee, PO Box 413, WI 53201, Milwaukee, USA}

\author{Kovi Rose}
\affiliation{Sydney Institute for Astronomy, School of Physics, The University of Sydney, NSW 2006, Australia}
\alsoaffiliation{Australia Telescope National Facility, CSIRO Space and Astronomy, PO Box 76, Epping, NSW 1710, Australia}

\author{Andrew Zic}
\affiliation{Australia Telescope National Facility, CSIRO Space and Astronomy, PO Box 76, Epping, NSW 1710, Australia}
\alsoaffiliation{ARC Centre of Excellence for Gravitational Wave Discovery (OzGrav), Australia}

\received {dd Mmm YYYY}
\revised  {dd Mmm YYYY}
\accepted {dd Mmm YYYY}
\published{22 September 202X}


\begin{document}

\begin{abstract}
We report the discovery of periodic radio bursts from the T8 brown dwarf \bdfull\ (\bdshort), detected with ASKAP and MeerKAT. The emission is characterised by a stable periodicity of $1.149 \pm 0.002$ hr, placing \bdshort\ among the fastest rotating radio-detected ultracool dwarfs (UCDs). The radio bursts are broadband and highly variable, reaching a peak spectral luminosity of $L_{\nu} = 5.6 \times 10^{15}\,\mathrm{erg\,s^{-1}\,Hz^{-1}}$, making \bdshort\ the brightest periodic UCD to date. The high brightness temperature of the radio bursts ($T_B>10^{11}$ K), combined with the rotational modulation, strongly favour an auroral electron cyclotron maser emission (ECME) origin. In contrast to most known radio-bursting UCDs, the bursts exhibit relatively low and variable circular polarisation fractions (typically $\sim 20\%$). These properties suggest significant depolarisation of intrinsically highly polarised ECME, likely due to propagation effects in a structured magnetospheric plasma. Notably, we also observe a polarisation reversal within a single burst profile, which is currently unexplained. We model the emission using a dipolar magnetic field with emission arising from a single active field line, and find a moderate inclination angle of $i = 47.9^{+11.4}_{-9.2}$ degrees and a magnetic obliquity of $\beta = 40.1^{+9.2}_{-10.9}$ degrees.
\bdshort\ expands the parameter space of radio-active UCDs toward higher luminosities and lower polarisation fractions, suggesting that current surveys may be sensitive to a larger population than previously recognised, while searches relying on methods that assume high circular polarisation fractions risk missing a subset of UCDs. More broadly, this discovery reinforces the idea that rapid rotation plays a key role in generating strong dipolar magnetic fields in UCDs that can give rise to auroral radio emission.
\end{abstract}

\section{Introduction}

Ultracool dwarfs (UCDs) occupy the regime between low-mass stars and giant exoplanets \citep{burrows2001theory}. They are broadly defined as objects with spectral type M7 and later, including L- T- and Y-dwarfs. The interiors of UCDs are fully convective, making them key laboratories for testing stellar dynamo theory. Typically, stellar magnetic fields of stars like the Sun are thought to be generated by a dynamo process that depends on the shearing motions at the radiative-convective transition zone \citep{brandenburg2005astrophysical}. An increasing number of UCDs have been shown to host strong, stable magnetic fields, capable of powering radio emission \citep[e.g.][]{pineda2017panchromatic, kao2018strongest}. This calls for a new dynamo process model in UCDs. These fully convective conditions are analogous to those expected in giant exoplanet atmospheres. Studying the magnetic field in UCDs can help us understand magnetic field generation in exoplanets \citep[see e.g.][]{nichols2012origin, showman2020atmospheric}.

Radio observations have proven to be a powerful diagnostic of magnetism in UCDs. Two distinct emission mechanisms are typically observed: (incoherent) gyrosynchrotron radiation and (coherent) electron cyclotron maser emission (ECME), see e.g. \cite{callingham2024radio}. Coherent plasma emission is usually not considered, because UCDs are unable to sustain a highly ionised plasma in their corona, which is a prerequisite for plasma emission at radio frequencies \citep{dulk1985radio}. For plasma‑like behaviour to arise, electromagnetic interactions must dominate over neutral collisions. This condition is not satisfied in the cool, predominantly neutral atmospheres of late-type UCDs. Thermal ionisation models indicate that while M-dwarfs may still reach ionisation fractions of $10^{-1}$, these values decline rapidly with temperature, dropping to $10^{-4} - 10^{-3}$ for objects near 1000 K \citep{rodriguez2015reference}. These ionisation levels are insufficient to maintain the plasma densities necessary for efficient plasma emission. Moreover, any dense plasma that does form would likely attenuate the radio emission via free-free absorption, further inhibiting observable emission. 

Gyrosynchrotron emission often appears as faint quiescent (unpulsed) emission that, unlike the highly polarised ECME bursts, is typically only weakly polarised. Recent VLBI imaging of LSR~J1835$+$3259 resolved this quiescent component into a stable, double-lobed structure closely resembling Jupiter's radiation belts, providing the first direct evidence that such weakly polarised emission traces magnetically trapped plasma rather than auroral beaming \citep{climent2023evidence,kao2023resolved}. In contrast, ECME produces highly circularly polarised bursts, often periodic with the rotation period of the star \citep[e.g.][]{hallinan2008confirmation}. ECME is now widely interpreted as auroral in nature, analogous to the magnetospheric radio emission of Jupiter \citep{zarka1998auroral}, and provides a direct measurement of the local magnetic field strength through the cyclotron frequency \citep{hallinan2007periodic,hallinan2008confirmation}. Some systems show both quiescent and bursting emission, hinting at the presence of both emission processes.

Radio emission from brown dwarfs was first detected from the M9 dwarf LP 944$-$20, which showed radio bursts with moderate circular polarisation ($\sim30\%$) attributed to synchrotron emission; this emission did not follow a strict periodicity \citep{berger2001discovery}. The discovery of strictly periodic, highly circularly polarised radio bursts from the M9 dwarf TVLM 513$-$46546 proved a breakthrough for the field \citep{hallinan2007periodic}. Its bursts reach 100\% circular polarisation and brightness temperatures exceeding $2\times10^{11}$ K, firmly establishing ECME, rather than gyrosynchrotron radiation, as the responsible emission mechanism \citep{hallinan2008confirmation}. Periodic radio emission has since been confirmed in a dozen UCDs spanning late-M, L and T spectral types. Every source in this sample shows radio bursts with circular polarisation fractions exceeding $50\%$, making ECME the most likely emission mechanism throughout. We review the individual properties of these systems in Section~\ref{sec:UCD_context} and construct a sample in Table~\ref{tab:periodic_UCDs}.

Since only a dozen UCDs with periodic radio emission have been detected, each new detection provides valuable insight into the broader UCD population. Most periodic radio UCDs known to date have been identified through targeted studies of nearby UCDs or untargeted searches that prioritise sources with strong, highly circularly polarised emission \citep[see e.g.][]{rose2023periodic}, a strategy well-matched to bright aurorae but one that risks leaving periodic emitters with weaker or less polarised signatures undetected. Untargeted, wide-field surveys open up a complementary discovery space, capable of uncovering periodic UCDs independent of their polarisation properties and so building a more complete, less biased census of the population.

\subsection{\bdfull}
In this paper, we report on the radio observations of \askapfull\ (\askapshort\ hereafter), which was discovered as a highly variable radio source in the Variables and Slow Transients survey \citep[VAST;][]{murphy2013vast} Extragalactic DR 1 \citep{deruiter2026askap}. We identify the T8 dwarf \bdfull\ (\bdshort\ hereafter) as the multi-wavelength counterpart to \askapshort.

We present the radio and multi-wavelength observations of \bdshort\ in Section~\ref{sec:observations}. After the discovery of \bdshort\ in ASKAP observations, we performed follow-up with ATCA and MeerKAT to characterise \bdshort\ fully. In Section~\ref{sec:results}, we calculate the brightness temperature and luminosity, characterise the polarisation properties and determine the radio period. In Section~\ref{sec:discussion} we use those properties to establish the emission mechanism, we infer the minimum magnetic field strength of \bdshort\ and perform detailed analysis of the radio light curve to constrain system properties. In Section~\ref{sec:UCD_context} we construct a sample of periodically emitting UCDs to put \bdshort\ into context; its low circular polarisation and high radio luminosity are outliers compared to the sample (Figure~\ref{fig:pol_lum}), and its rapid rotation strengthens the hypothesis that rapid rotation plays a key
role in generating strong dipolar magnetic fields in convective dynamos \citep{kao2018strongest}. Finally, we summarise our conclusions in Section~\ref{sec:conclusion}.

\section{Observations} \label{sec:observations}
\subsection{ASKAP} \label{sec:obs_askap}
\begin{figure*}
    \centering
    \includegraphics[width=\linewidth]{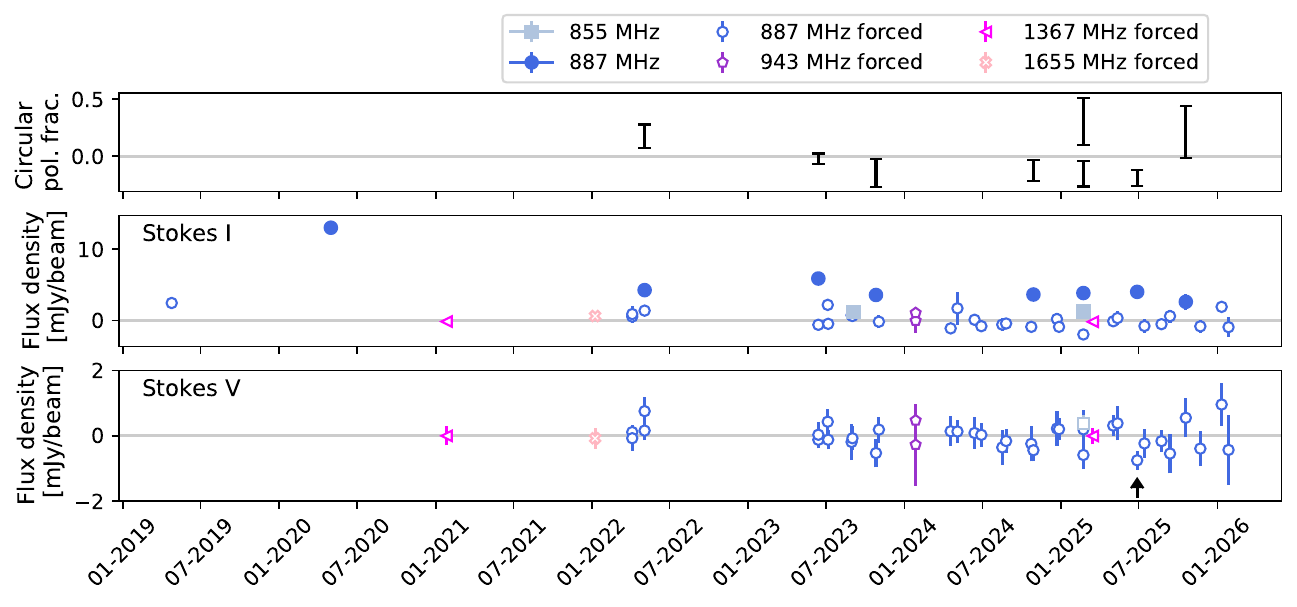}
    \caption{ASKAP light curve of \askapshort. Solid markers indicate $> 5\sigma$ detections from \textsc{Selavy} in the \textsc{ASKAPSoft} pipeline, while open markers show forced flux densities obtained by fitting a 2D Gaussian at the source position in case of non-detections (see text for details). The middle panel shows the Stokes I integrated flux density, whereas the bottom panel shows the Stokes V integrated flux density. Marker colour/shape denote observing frequency. The top panel shows the circular polarisation fraction for the $> 5\sigma$ Stokes I detections. The arrow in the bottom panel shows the only ASKAP observation with a tentative circular polarisation detection (see the rightmost panel of Figure \ref{fig:examples_VAST}).}
    \label{fig:askap_lightcurve}
\end{figure*}

\askapshort\ was detected as a highly variable source in the first VAST Extragalactic data release \citep{deruiter2026askap}. VAST is one of the key Survey Science Projects running on the Australian SKA Pathfinder \citep[ASKAP;][]{johnston2008science, hotan2021australian}.  Figure \ref{fig:askap_lightcurve} shows all archival ASKAP observations of \askapshort\ up to April 2026. 
The flux densities were measured by the \textsc{Selavy} source finding software \citep{whiting2012duchamp}, incorporated into the \textsc{ASKAPSoft} pipeline. A measurement was considered a detection if the peak flux density exceeded 5 times the local root mean square (rms) noise, indicated by solid markers in Figure \ref{fig:askap_lightcurve}. Additionally, Figure \ref{fig:askap_lightcurve} shows forced flux density measurements as open markers. These were obtained by fitting a two‑dimensional Gaussian at the average source position in any observation lacking a $>5\sigma$ detection. The shape and orientation of the Gaussian were fixed to match the point spread function of the observation, and only the peak amplitude was allowed to vary. These fitted flux densities could be negative due to the nature of radio images (close to zero-mean noise). A negative Stokes I forced flux implies that the measurement is consistent with noise. Note that a negative Stokes V flux measurement does have a physical meaning.  We follow the IAU/IEEE convention \citep{hamaker1996understanding} where a positive Stokes V flux corresponds to right-hand circular polarisation and a negative Stokes V flux to left-hand circular polarisation. The top panel in Figure \ref{fig:askap_lightcurve} shows the Stokes I flux, whereas the bottom panel shows the Stokes V flux. 

The different frequencies of the ASKAP observations are indicated by the marker colour/shape. The majority of the observations were obtained as part of the VAST survey (887 MHz) with typical integration times of 12 minutes. Some observations at 887 MHz, as well as the 943, 1367 and 1655 MHz observations were collected by the RACS survey \citep{mcconnell2020rapid, duchesne2024rapid, duchesne2025rapid}, with observing durations of 15 minutes. Finally, the two 855 MHz detections were obtained as part of the FLASH survey \citep{allison2022first}, with integration times of two hours. There was only one observation (indicated with the black arrow in Figure \ref{fig:askap_lightcurve}) with a significant (negative) Stokes V detection. We find no archival radio detections when querying the \bdshort\ coordinates in Vizier \citep{vizier}. Just considering the 887 MHz observations done by RACS and VAST, we find that \askapshort\ is detected 7 times in 36 observations, indicating a duty cycle of $\sim20\%$.\\

\begin{figure}
    \centering
    \includegraphics[width=\linewidth]{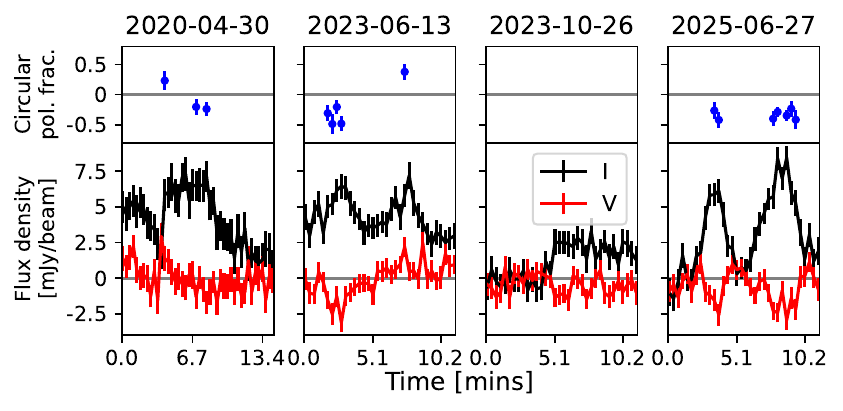}
    \caption{Four exemplary VAST detections showing the Stokes I (black) and Stokes V (red) evolution within one VAST observation in the bottom panel, and the circular polarisation fraction in the top panel. The rightmost panel is the VAST observation with the strongest evidence for circular polarisation (indicated with a black arrow in Figure \ref{fig:askap_lightcurve}).}
    \label{fig:examples_VAST}
\end{figure}

Figure \ref{fig:examples_VAST} shows four examples of VAST detections of \askapshort. We extracted the Stokes I and Stokes V dynamic spectrum from the measurement sets using \textsc{DStools}\footnote{\url{https://github.com/askap-vast/dstools}} \citep{Pritchard2025a}. We did not apply additional flagging and averaged the time resolution by a factor of two to $20$ seconds for all observations. We only show the light curves in Figure \ref{fig:examples_VAST}, as there was no frequency structure across the 288 MHz ASKAP bandwidth; the bursts are broadband. Figure \ref{fig:examples_VAST} presents the Stokes I (black) and Stokes V (red) evolution within one VAST observation in the bottom panel, and the circular polarisation fraction in the top panel. We only included circular polarisation fractions if the absolute value of the Stokes V flux is $>1.25$ mJy/beam, corresponding to five times the typical noise of $0.25$ mJy/beam. The rightmost panel corresponds to the VAST observation with the strongest evidence for circular polarisation (indicated with a black arrow in Figure \ref{fig:askap_lightcurve}).

\subsubsection{EMU}\label{sec:obs_EMU}
We analysed two observations from the EMU survey \citep{norris2011emu}, corresponding to the $\rm{EMU}~1502+04$ pointing. This pointing was observed twice for five hours under scheduling block IDs SB81088 and SB81400, on 2026-01-12 and 2026-01-24, respectively, at a central frequency of 943 MHz with a 288 MHz bandwidth. We re-imaged these measurement sets by selecting phased array feed beam 23, centred at (RA, Dec) = ($228.1570^{\circ}$, $5.0591^{\circ}$). We increased the image size beyond the typical $\sim1~\rm{deg}^2$ to include the location of \askapshort, as \askapshort\ lies just outside the typical $\sim30~\rm{deg}^2$ EMU field of view. We note that this is also the reason why these EMU observations do not appear in Figure \ref{fig:askap_lightcurve}. Using \textsc{DStools}, we extracted the dynamic spectra for these observations.
Both observations suffered from substantial radio frequency interference (RFI) between 850–890 MHz, and we fully flagged this frequency range. For SB81088, we applied additional flagging between 940–960 MHz, and for SB81400 between 915–960 MHz. We averaged the time resolution to 60 seconds for both observations.
We detected faint bursts from \askapshort\ in both observations. There is marginal evidence for Stokes V in the folded burst profiles of SB81400 (see Figure \ref{fig:EMU_periodogram} and Section \ref{sec:results_periodicity}).

To constrain the periodicity of \bdshort, understand the spectral properties of the emission, and carry out a detailed analysis of its polarisation properties, we perform follow-up observations with ATCA and MeerKAT.

\begin{figure*}
    \centering
    \includegraphics[width=0.9\linewidth]{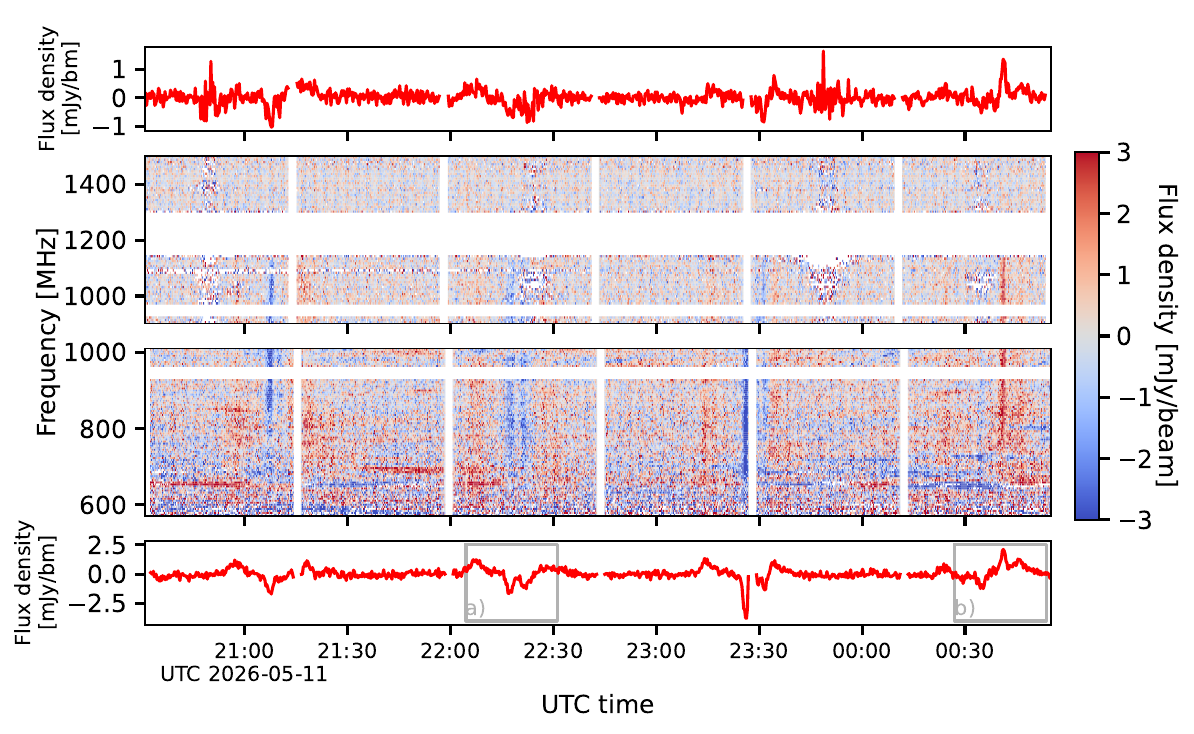}
    \caption{Stokes V dynamic spectrum (570-1500 MHz) from the MeerKAT observation of \askapshort. The upper two panels show the dynamic spectrum and light curve for the L-band observations; the lower two panels show the same for the UHF-band. Horizontal gaps correspond to frequencies that were flagged due to radio frequency interference (see text) and the vertical gaps correspond to phase calibration scans. The horizontal features visible around 650 MHz are attributable to residual radio frequency interference (RFI).}
    \label{fig:MKT_DS_V}
\end{figure*}

\subsection{ATCA} \label{sec:obs_atca}
We observed \askapshort\ with ATCA on 2025-11-12 from 00:30 to 05:30 UTC. ATCA was in the 6A extended configuration, and the observations used the L-band (1.1-3.1 GHz). The near-zero declination of \askapshort\ resulted in a highly elongated synthesised beam. The standard primary flux calibrator J1939$-$6342 was observed at the start of the observation, and a phase calibrator was observed every 30 minutes. We initially observed J1518$+$045 as a phase calibrator, but it was not used in the final calibration. The second phase calibrator, J1548$+$056, exhibited more favourable structure and closure phases (i.e. a more isolated and compact bright source), and was therefore used for phase calibration.

We reduced the data using a standard workflow in the Common Astronomy Software Applications package \citep[\textsc{CASA};][]{casa2022casa}. In addition, we applied self-calibration to the phase solutions to improve the calibration quality. Automated flagging routines (\texttt{tfcrop} and \texttt{rflag}) were applied at multiple stages of the calibration process, as well as to the final calibrated target data. Imaging was performed using \textsc{WSCLEAN} \citep{offringa2014wsclean}, with Briggs weighting (robust parameter 0.5), auto-threshold 3, auto-mask 5, mgain 0.8, and \texttt{niter} = 30,000. We did not detect \askapshort; the rms noise at the target position is $0.03$ mJy beam$^{-1}$.

We further used \textsc{DStools} to extract a dynamic spectrum to search for short-duration bursts that may have been averaged out in the integrated image. Frequencies below 1300 MHz were excluded due to severe radio frequency interference (RFI), and additional manual flagging was applied to the 1940–1970 MHz range. We explored a range of time and frequency averaging schemes, as well as different upper frequency cutoffs. The standard deviation of the imaginary component of the complex visibilities was used to estimate the noise level in the Stokes I light curves. We detected a single burst at approximately the $3\sigma$ level, depending on the averaging scheme. We consider this burst unlikely to originate from \askapshort: it occurs immediately prior to time steps flagged as RFI by \texttt{tfcrop}, and a similar feature is present in the imaginary component, suggesting a non-astrophysical origin. Finally, we note that the non-detection in these ATCA observations is likely due to the upper cutoff frequency of $\sim$1300 MHz inferred from subsequent MeerKAT observations (see Figure \ref{fig:spectral_shape}).

\subsection{MeerKAT} \label{sec:obs_mkt}

We observed \askapshort\ with MeerKAT \citep{jonas2016meerkat} under DDT-20251022-ID-01 on 2026-05-11 from 20:03 to 2026-05-12 01:07 UTC.
The array was configured into two sub-arrays, enabling simultaneous observations of \askapshort\ in the UHF band (544–1088 MHz) and L-band (856–1712 MHz). The data were recorded with a time resolution of 2 seconds. We used J1939$-$6342 as the primary flux and bandpass calibrator, J1331$+$3030 (3C286) as the polarisation calibrator, and J1550$+$0527 for phase calibration. The phase calibrator was observed every 45 minutes, resulting in a total on-source time of 4.2 hours. The data were reduced using the \textsc{polkat} pipeline\footnote{\url{https://github.com/AKHughes1994/polkat/tree/polkat_QC_selfcal}} \citep{hughes2025polkat}. We note that the observations presented here are repeat observations, as earlier data obtained under this programme proved impossible to calibrate reliably. Those initial observations were conducted on 2025-11-09 from 07:25 to 12:33 UTC, coinciding with the peak (07:35 UTC) of an X1.7 solar flare event \citep{yasyukevich202612}. 

We detect \askapshort\ in the integrated images, with peak flux densities of $1.48 \pm 0.02$ mJy/beam and $0.20 \pm 0.01$ mJy/beam in the UHF and L-bands, respectively. Figure \ref{fig:MKT_DS_V} shows the Stokes V dynamic spectrum of this observation. The lower two panels correspond to the UHF-band dynamic spectrum and light curve, while the upper two panels show the equivalent products for L-band. There is a small time offset between the observation start (and calibrator scans) in UHF and L-band. In Figure \ref{fig:MKT_DS_V} the time resolution was averaged down to 16 seconds, and the frequency resolution is 5.39 MHz (UHF) and 8.43 MHz (L-band).

We applied manual flagging to the frequency ranges 925–965 MHz and 1145–1300 MHz, and removed the noisy portion of the L-band above 1500 MHz. We also used the imaginary component of the complex visibilities for further RFI mitigation, flagging time–frequency pixels with outlier values in the imaginary dynamic spectrum. Some residual RFI nonetheless remains as horizontal streaks around 650 MHz, matching persistent but faint digital TV interference\footnote{\url{https://skaafrica.atlassian.net/wiki/spaces/ESDKB/pages/305332225/Radio+Frequency+Interference+RFI}}. These streaks are visible in all polarisations and in the imaginary component of the complex visibilities, confirming the RFI nature, but are too faint to exceed the significance threshold used for the imaginary-component dynamic spectrum flagging.

\begin{figure}
    \centering
    \includegraphics[width=\linewidth]{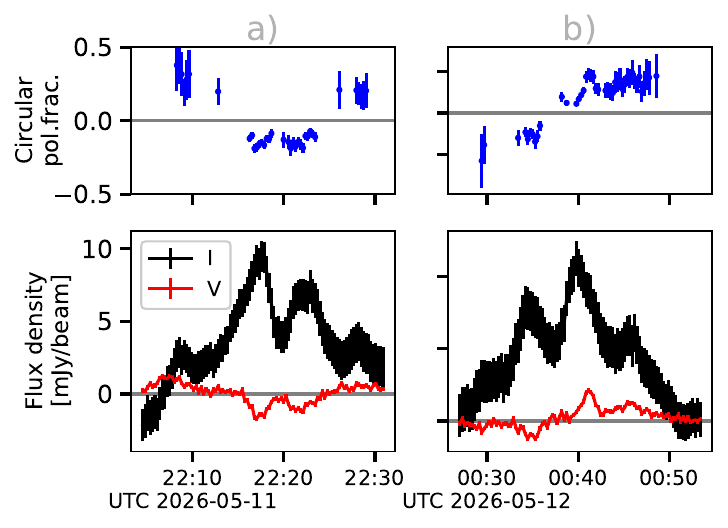}
    \caption{Detailed MeerKAT UHF light curves of two of the four Stokes V bursts (as indicated in the bottom panel in Figure \ref{fig:MKT_DS_V}). These figures show the Stokes I (black) and Stokes V (red) evolution in the bottom panel, and the circular polarisation fraction in the top panel.}
    \label{fig:MKT_DS_zooms}
\end{figure}

\begin{figure}
    \centering
    \includegraphics[width=\linewidth]{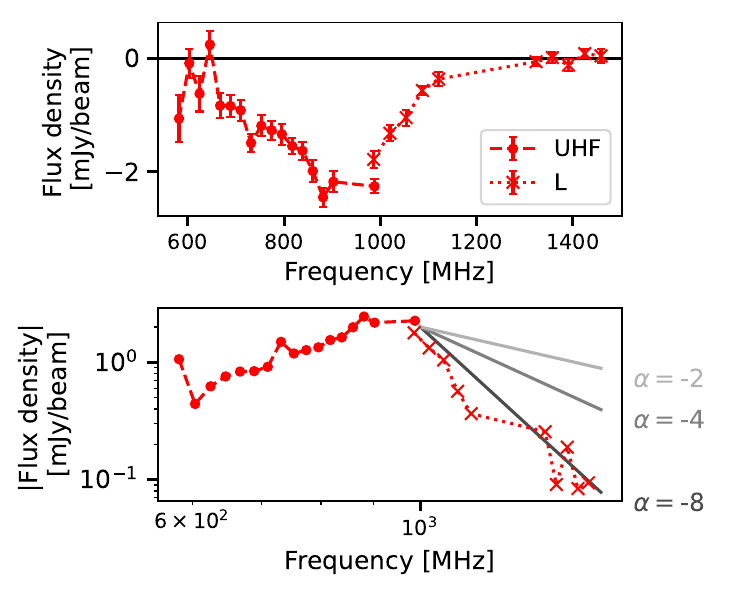}
    \caption{Average spectral shape of the first two bursts in the MeerKAT observations (see Figure \ref{fig:MKT_DS_V}). The spectrum has been constructed by averaging over 1.5 minutes in time, centred on the time when the light curve peaks. The top panel shows the Stokes V flux density (including $1\sigma$ error bars) as a function of frequency. The bottom panel shows the absolute Stokes V flux, as a function of frequency in log-space, with different spectral indices overplotted.}
    \label{fig:spectral_shape}
\end{figure}

We do not show the Stokes I dynamic spectrum because it is strongly affected by artefacts arising from bright off-axis emission. In particular, sources with flux densities of $0.3$ and $0.7$ Jy located $1.5$ and $1.4$ degrees from the phase centre, respectively (that is, beyond the FWHM of the primary beam), produce artefacts across the deep image. When subtracting this deep image from the visibilities to create a dynamic spectrum, these imperfections manifest as residual features. Although the severity of these artefacts is reduced through self-calibration, they cannot be completely removed. Self-calibration is performed within the \textsc{polkat} pipeline using \textsc{QuartiCal} \citep{Quartical2023}. Instead, we focus on two bursts for which reliable Stokes I light curves could be extracted, and present these alongside the Stokes V light curves in Figure \ref{fig:MKT_DS_zooms}. As in Figure \ref{fig:examples_VAST}, the black lines show the Stokes I flux, the red lines show the Stokes V flux, and the upper panels display the circular polarisation fraction for significant Stokes V detections ($>0.5$ mJy/beam).
The light curves and dynamic spectra show clear periodic behaviour. The Stokes V bursts are predominantly negative and double-peaked, except for the final burst, where the second peak exhibits positive Stokes V. Small enhancements in Stokes V flux are seen immediately before and after the main burst, particularly in the UHF light curve. The detailed light curves (Figure \ref{fig:MKT_DS_zooms}) confirm moderate circular polarisation fractions, consistent with the ASKAP observations. Focusing on the peaks of the Stokes V emission, we measure polarisation fractions of $-0.19 \pm 0.03$ and $-0.19 \pm 0.05$ for panel a, and $-0.17 \pm 0.04$ and $0.23 \pm 0.04$ for panel b. The dynamic spectrum reveals a clear upper cutoff frequency in the flagged bandwidth range between 1145 and 1300 MHz. Figure \ref{fig:spectral_shape} shows the average spectral shape of the first two bursts. A corresponding lower cutoff frequency is less evident. The bursts appear to extend to lower frequencies in the second and third bursts, but this may reflect their increasing brightness. 

Finally, we use the MeerKAT UHF observations to obtain an upper limit on the quiescent emission of \askapshort. From the 4.2 hours of on-source integration time, roughly 2.5 hours are off-pulse. Scaling the rms noise in the deep image ($50 \,\mu\rm{Jy/beam}$) to the off-pulse time ($\times \sqrt{4.2/2.5}$) yields a $65\,\mu\rm{Jy/beam}$ upper limit on the quiescent emission in UHF-band.

\subsection{Identification as a T8 dwarf}
\begin{figure}
    \centering
    \includegraphics[width=\linewidth]{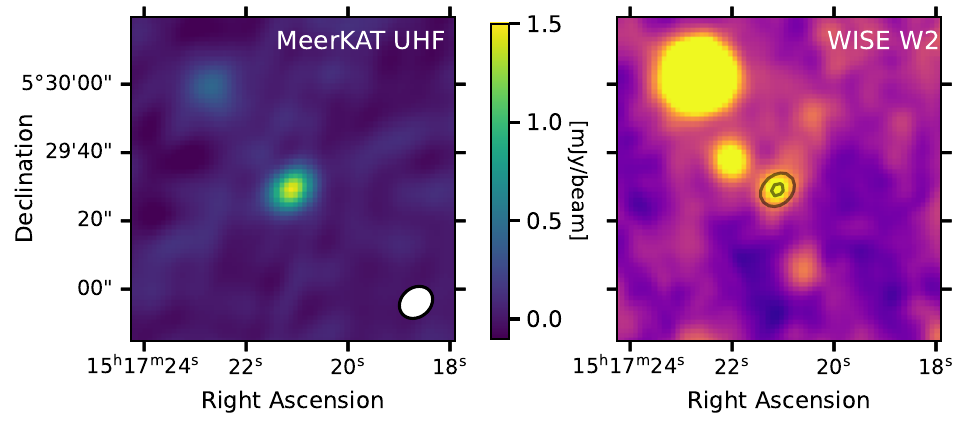}
    \caption{Left: MeerKAT UHF Stokes I detection of \askapshort, observed on 2026-05-11 (DDT-20251022-ID-01), imaged with uniform weighting to improve the astrometric accuracy. The ellipse in the lower right corner of the radio image shows the FWHM of the restoring beam. The MeerKAT position has been corrected according to the proper motion of \bdshort\ (see text), such that the radio detection is corrected to the time of the WISE observation. Right: WISE W2 detection of \bdshort, observed on 2010-02-06. The MeerKAT Stokes I contours at 50\% and 90\% of the peak flux density are overlaid.}
    \label{fig:mkt_wise_cutouts}
\end{figure}

We cross-matched the radio source \askapshort\ with the infrared source \bdfull, which has Wide-field Infrared Survey Explorer \citep[WISE;][]{wright2010wide} magnitudes of $W1 > 18.10$, $W2 = 15.13 \pm 0.08$, $W3 > 13.13$, and $W4 > 9.52$. Figure \ref{fig:mkt_wise_cutouts} shows the MeerKAT UHF Stokes I detection of \askapshort\ (see Section \ref{sec:obs_mkt}) in the left panel, while the right panel shows the corresponding WISE W2 cutout, with grey contours indicating the radio emission. 

\bdshort\ has a proper motion of $\mathrm{pm}{\mathrm{RA}} = -60.3 \pm 2.0$ mas yr$^{-1}$ and $\mathrm{pm}{\mathrm{Dec}} = 189.2 \pm 2.3$ mas yr$^{-1}$ \citep{kirkpatrick2021field}. Before proper motion correction, \askapshort\ and \bdshort\ are  $3.2\arcsec$ offset. After correcting for the proper motion of \bdshort\ over the 16-year baseline between the WISE and MeerKAT observations, we find only a $0.7\arcsec$ offset, leading us to conclude that \askapshort\ is \bdshort.
Figure \ref{fig:mkt_wise_cutouts} shows the MeerKAT observation proper motion corrected to the time of the WISE observations. 

\bdshort\ was identified as a T dwarf using WISE photometry by \cite{davy2012further}. 
\cite{mace2013study} determined a spectral type of T8 for \bdshort\ using the Near-Infrared Spectrometer \citep[NIRSPEC;][]{mclean1998design, mclean2000performance} on the 10 m W. M. Keck Observatory. This was independently confirmed by \cite{burningham201376}, who classified the source as T8p (peculiar) using the Gemini Near Infrared Spectrograph \citep[GNIRS;][]{elias2006proc} on the Gemini North Telescope.
\cite{kirkpatrick2021field} presented a census of 525 nearby L, T, and Y dwarfs based on \textit{Spitzer} observations and derived a distance to \bdshort\ of $21.7 \pm 0.1$ pc. Although they did not determine an effective temperature for \bdshort, the temperatures of other T8 dwarfs in their sample range from $690 \pm 80$ K to $820 \pm 80$ K.

\section{Results} \label{sec:results}
\subsection{Radio luminosity and brightness temperature}\label{sec:results_luminostiy_brightness_temp}

We calculate the luminosity of the radio emission using the MeerKAT Stokes I peak flux densities of $\sim10$ mJy/beam at 816 MHz. This flux estimate comes from the two bursts with reliable Stokes I dynamic spectra\footnote{From 10-minute snapshots, we determine that the Stokes I flux for all MeerKAT bursts is very similar; between 4 and 6 mJy/beam in the 10-minute integration covering the peak of the burst.}, presented in Figure \ref{fig:MKT_DS_zooms}. \bdshort\ is at a distance of 21.7 pc. Converting this to a spectral luminosity yields $L_{\nu}=5.6 \cdot 10^{15} \, \rm{erg}\,\rm{s}^{-1}\,\rm{Hz}^{-1}$. 

The only unknown in the brightness-temperature calculation is the size of the emission region, so to obtain a lower limit we assume the largest plausible region, which for gyrosynchrotron radiation is set by the scale of the stellar magnetic field and is therefore on the order of the stellar radius \citep{gudel2002stellar, berger2006radio}. The radius of an ultracool dwarf lies between 0.85 and 1.0 $R_{Jup}$ \citep{baraffe2003evolutionary, zhang2021uniform}.

Assuming that the emission region has a maximum size of three stellar radii, with a stellar radius of $1.0~R_{\rm Jup}$, we obtain a deliberately conservative brightness-temperature lower limit of $T_B > (2.7$--$16) \times 10^{11}$~K for flux densities of $2$--$10$~mJy. This source size is intentionally generous, as it maximises the emitting area and therefore minimises the inferred brightness temperature. Even under this assumption, the brightness temperature exceeds the gyrosynchrotron brightness-temperature limit of $T_B \sim 10^{10}$--$10^{11}$~K, implying a coherent emission mechanism instead \citep{dulk1985radio}. For ECME, the emission region is much smaller ($10^{-1}$ to $10^{-3}$ times the stellar radius \citep{zarka1998auroral, melrose1982electron, treumann2006electron}), and thus the inferred brightness temperature would be well above $10^{14}$ K.

\subsection{Polarisation} \label{sec:results_polarisation}
We summarise the polarisation properties of \bdshort\ as follows:
\begin{itemize}
    \item Most of the ASKAP bursts do not show a significant circular polarisation fraction. 
    The ASKAP detections on 2023-06-13 and 2025-06-27 have a maximum circular polarisation fraction of $(-)50\%$ and $(-)40\%$ respectively (see Figure \ref{fig:examples_VAST}). We note that for slightly fainter bursts, with peak Stokes I flux of 5 mJy/beam, a circular polarisation fraction of ($+/-$)25\% falls below the rms noise in the Stokes V dynamic spectra.
    The circular polarisation fraction of the MeerKAT bursts is 20\%. Together with the ASKAP bursts, this indicates that the circular polarisation fraction is variable over time.
    \item The burst profile in the MeerKAT dynamic spectrum consists of four components. The main bursts are double-peaked and highly polarised (negative Stokes V). Additionally, there is a faint pre- and post-burst peak, with opposite and lower polarisation fractions. 
    \item Although most bursts show a similar structure, one burst in the MeerKAT observation clearly deviates (see panel b, in Figure \ref{fig:MKT_DS_zooms}). This burst appears at the same phase as the previous `main' bursts; however, the polarisation of the double-peaked structure flips. The first peak of the main bursts shows negative Stokes V, like all other bursts, but the second peak in the main burst shows positive Stokes V. The pre- and post-main bursts' polarisation properties seem similar to all other bursts. There is one other hint of this polarisation change in the 2023-06-13 ASKAP observation (see Figure \ref{fig:examples_VAST}), but the light curves are at low signal-to-noise. 
\end{itemize}

These polarisation properties of the radio emission are atypical for ECME. Below, we discuss the polarisation properties for gyrosynchrotron emission and ECME.\\

Gyrosynchrotron emission produces only moderate circular polarisation, typically well below 100\%, with the exact fraction depending strongly on viewing geometry, harmonic number, and magnetic field structure \citep{dulk1985radio}.  Solar observations of bursts from gyrosynchrotron emission typically show moderate circular polarisation, commonly of order $\sim10 - 20\%$, with higher values occurring only under favourable conditions such as strong, well-ordered magnetic fields and specific viewing geometries \citep{bastian1998radio,zhu2021microwave}.\\

Although ECME is intrinsically highly circularly polarised, often approaching 100\% \citep{dulk1985radio, hallinan2008confirmation}, this polarisation can be substantially reduced during propagation through structured plasma. Mode coupling between the extraordinary (X) and ordinary (O) modes is most efficient in quasi-transverse regions, where the wavevector becomes nearly perpendicular to the magnetic field \citep{melrose1994reversal, treumann2006electron, shain2017role}; since the X- and O-modes carry opposite senses of circular polarisation, transferring even part of the power between them lowers the net observed polarisation fraction. 
More generally, linear mode coupling in inhomogeneous plasmas is well known to reduce and even reverse polarisation in solar radio emission \citep{dulk1980position}. Additional depolarisation arises from propagation through density gradients, where partial reflections and scattering progressively mix the two modes \citep[see e.g. Solar observations by][]{kaneda2017polarization}. The moderate circular polarisation fractions of the radio emission of \bdshort\ are consistent with either a gyrosynchrotron emission scenario or depolarised ECME. \\

The last burst in Figure \ref{fig:MKT_DS_V} (burst b), however, shows a sudden reversal in circular polarisation, present only in its secondary peak while the first peak remains negative. Such a flip is difficult to reconcile with a simple geometric picture, since it occurs at essentially the same rotational phase as the other bursts and is not accompanied by a distinct interpulse component; ECME models usually attribute opposite polarisation senses to emission from opposite magnetic hemispheres or to viewing different field orientations during rotation \citep[e.g.][]{hallinan2007periodic, hallinan2008confirmation}.

Neither an ECME nor a gyrosynchrotron origin is naturally expected to produce a polarity reversal at the same rotational phase, so propagation effects may instead be responsible. Radio waves can switch circular polarisation state if their path crosses a point where the line-of-sight field component changes sign. This happens for instance, where the ray grazes the magnetic equator or a current sheet. The outcome of this crossing is set by a transition frequency $f_t \propto (n_e B^3 L)^{1/4}$, where $n_e$ and $B$ are the density and field strength at the crossing and $L$ the local gradient scale length: below $f_t$ coupling is weak and the polarisation reverses; above it coupling is strong and the polarisation is preserved \citep{melrose1994reversal, treumann2006electron, shain2017role}. Scaling from solar values ($n_e\sim10^{10}\,\mathrm{cm^{-3}}$, $B\sim100$~G, $f_t\sim10$~GHz) to a plausible UCD magnetosphere ($n_e\sim10$--$100\,\mathrm{cm^{-3}}$; \cite{nichols2012origin}, comparable $B$) gives $f_t$ of order a few hundred MHz, below our observing band. Coupling should therefore typically be strong and the polarisation preserved, as seen in the three other MeerKAT (and ASKAP) bursts, with burst b's reversal instead requiring a transient sharpening of the local density gradient. This scaling is illustrative, given how poorly constrained $n_e$, $B$, and $L$ are in a UCD's outer magnetosphere, but it shows that a flip requires a specific combination of field and density conditions -- potentially explaining why such reversals are not seen more widely across the UCD population. Alternatively, the burst may originate from a distinct, transient emission site rather than the steady global field, which would likewise appear only occasionally and suggest that the field can depart locally from a simple dipole.

\subsection{Periodicity} \label{sec:results_periodicity}
\begin{figure}
    \centering
    \includegraphics[width=0.99\linewidth]{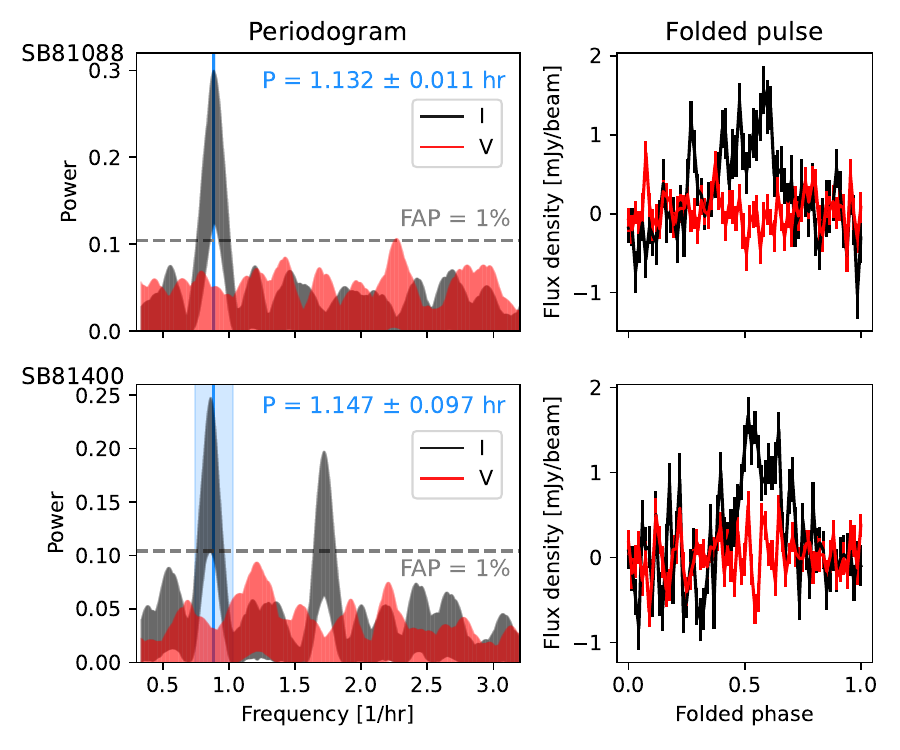}
    \caption{Bootstrapped representations of the Lomb-Scargle periodogram for the 5-hour EMU observation (left) and the folded burst profile for the most prominent period in the periodogram (right). The top row corresponds to SB81088, observed on 2026-01-12, while the bottom row corresponds to SB81400, observed on 2026-01-24. The black lines correspond to Stokes I power/flux, and the red lines to Stokes V. The horizontal grey dashed line in the periodogram shows the $1\%$ false alarm probability power. The blue vertical line indicates the most prominent frequency in the periodogram.}
    \label{fig:EMU_periodogram}
\end{figure}

As discussed in Section \ref{sec:obs_EMU}, we find repeat bursts of \bdshort\ in the EMU observations. Figure \ref{fig:EMU_periodogram} shows the periodogram and folded burst profile for these observations. The 5-hour observations each show a prominent peak at  
$1.132 \pm 0.011$ and $1.147 \pm 0.097$ hours in the Stokes I data, respectively. The periods (plus uncertainties) were estimated via bootstrapping. Each light curve was resampled 1000 times by perturbing the flux measurements according to their uncertainties, and the most prominent period was determined for each realization. The average of this distribution provides the most prominent period, whereas the standard deviation gives the $1\sigma$ uncertainty on the period.
The dashed horizontal grey lines show the 1\% false alarm probability (FAP). The second peak in the SB81400 periodogram is a harmonic frequency of the main peak. The right-side panels in Figure \ref{fig:EMU_periodogram} show the folded burst profiles when folding the data to the most prominent period. The black lines show the periodogram and folded burst profile for Stokes I emission, and the red lines correspond to Stokes V. We do not detect a significant periodicity in the EMU Stokes V data, which is likely due to the low circular polarisation fraction of the bursts.\\

\begin{figure}
    \centering
    \includegraphics[width=0.9\linewidth]{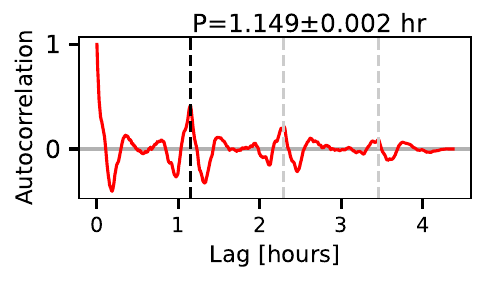}
    \caption{Auto-correlation function of the Stokes V MeerKAT UHF light curve. The black vertical line shows the period of \bdshort, the grey lines show the higher-order harmonics. Combining the three harmonics yields a period of $1.149\pm0.002$ hours.}
    \label{fig:MKT_ACF}
\end{figure}

The most reliable period estimate comes from the high-SNR bursts in the MeerKAT observations and is in agreement with the aforementioned periodicity estimates. Figure \ref{fig:MKT_ACF} shows the auto-correlation function for the Stokes V MeerKAT UHF light curve. The vertical lines indicate the harmonics of the period. We use all three harmonics to derive a period of $1.149\pm0.002$ hours. We use this as the radio period of \bdshort\ in the remainder of this work.

The temporal behaviour of radio emission provides a useful diagnostic of the underlying emission mechanism in UCDs \citep[see e.g.][]{callingham2024radio}. Gyrosynchrotron emission is expected to be stochastic or smoothly varying, as it originates from large populations of mildly relativistic electrons in extended, evolving coronal regions. Smooth variations in radio flux reflect changes in particle acceleration and plasma conditions \citep{dulk1985radio,bastian1998radio}.
In contrast, ECME is periodic by nature, as it is highly beamed, originating in compact regions tied to stable magnetic field geometries. As a UCD rotates, this produces lighthouse-like periodic bursts with stable phase and high contrast, analogous to planetary auroral radio emission \citep{zarka1998auroral,treumann2006electron,hallinan2007periodic}.
Thus, while gyrosynchrotron emission is typically stochastic, the presence of stable periodic bursts, as is the case for \bdshort, strongly favours an ECME origin.\\

We note that \cite{williams2015simultaneous} argued that the periodically varying, weakly to moderately circularly polarised (0--25\%) emission from 2MASS J13142039$+$1320011 can be explained by rotationally modulated gyrosynchrotron emission. However, the corresponding emission profile is broad in phase, whereas the bursts observed from \bdshort\ are considerably narrower relative to the rotation period.

\section{Discussion} \label{sec:discussion}

\subsection{Emission mechanisms}

Considering the brightness temperature, polarisation properties and periodicity of the radio emission from \bdshort\, as discussed in the previous sections, we conclude that ECME is the most likely emission mechanism. The inferred brightness temperature exceeds the expected gyrosynchrotron limit even under conservative assumptions, requiring a coherent emission process. In addition, the narrow, stable, and periodic bursts are naturally explained by the beamed emission geometry of ECME. In Section \ref{sec:results_polarisation} we explain how the intrinsically highly circularly polarised ECME can be depolarised to the circular polarisation fractions ($\sim 20\%$) observed in \bdshort. 

\subsection{Minimum magnetic field strength}
Periodic ECME in UCDs directly traces kilogauss-level magnetic fields inferred from the emission frequency \citep{kao2023resolved}. Based on the MeerKAT dynamic spectrum (Figure \ref{fig:MKT_DS_V}) and the average burst spectrum (Figure \ref{fig:spectral_shape}), we conservatively estimate the upper cutoff frequency of the ECME emission at 1.3 GHz. The minimum magnetic field strength in the emission region is $B=\nu_{\rm{cutoff}}/2.8=1.3/2.8=0.46$ kG \citep{dulk1985radio}. ECME operates at different harmonics of the cyclotron frequency. If we observe the first harmonic instead of the fundamental, the magnetic field strength would be $B=1.3\cdot0.5/2.8=0.23$ kG instead. \\

There are several reasons why this simple interpretation of the cutoff frequency may be oversimplified. First, the observed cutoff may not be intrinsic to the emission mechanism but instead result from absorption along the propagation path. Regions of stronger magnetic field may be present and capable of producing higher-frequency emission, but this emission could be attenuated by processes such as free-free absorption before escaping the magnetosphere. Second, ECME requires the plasma frequency ($\nu_p \propto \sqrt{n_e}$) to be significantly lower than the cyclotron frequency ($\nu_c \propto B$), i.e. $\nu_p \ll \nu_c$ \citep{dulk1985radio}. Consequently, the highest observable ECME frequency may be limited not only by the magnetic field strength but also by local plasma conditions, including the electron density. Finally, ECME is intrinsically highly directional, with radiation confined to narrow beams. As a result, we detect emission only when the beaming geometry is favourable. Higher-frequency emission from regions of stronger magnetic field may therefore exist but remain undetected if the corresponding emission beams do not intersect our line of sight.

\bdshort\ is one of the latest spectral type UCDs observed to produce radio emission, with the other being the T8 dwarf WISE~J062309.94$-$045624.6 \citep{rose2023periodic}, for which the magnetospheric properties were similarly constrained using radio bursts. For WISE~J062309.94$-$045624.6 the minimum magnetic field strength was estimated to be 0.71~kG (or 0.35 kG if emission is at the first harmonic). \\

Recent work by \cite{yiu2026magnetic} points out that it can be difficult to distinguish between the spectral evolution of the radio emission across the band (accounting for the sensitivity of the telescope) and an actual spectral cutoff. We present the average spectrum of the first two MeerKAT bursts in Figure \ref{fig:spectral_shape}. This spectrum is obtained by averaging over the peak time bin and the 6 surrounding time bins ($\sim$ 1.5 min). The top panel shows the Stokes V flux in the UHF and L band, whereas the bottom panel shows the absolute value of the Stokes V flux as a function of frequency, with logarithmic axes. We overplot in shades of grey different spectral indices. This shows that the spectral shape of the \bdshort\ bursts is characterised by a spectral cutoff, since the spectral index is around $-8$. \citet{yiu2026magnetic} argue that a steep spectral index ($\alpha\approx -4$ assuming a flux of 160 $\mu$Jy at 168 MHz, and an upper limit of $<16$ $\mu$Jy at 300 MHz) can only be explained by a spectral cutoff. Finally, we note the spectral drop-off towards the bottom of the frequency band. The bottom part of the UHF band is noisier (and still includes residual RFI, as indicated in Section \ref{sec:obs_mkt}); therefore, it is unclear whether the radio emission from \bdshort\ has a lower-frequency cutoff. From the bottom panel of Figure \ref{fig:spectral_shape}, the spectral shape at lower frequencies seems inconsistent with a spectral cutoff. Future observations with low-frequency instruments are required to accurately determine whether this is a low-frequency cutoff or the spectral evolution of the emission. 

\subsection{Active field line modelling} \label{sec:discussion_afl}
\begin{figure}
    \centering
    \includegraphics[width=0.99\linewidth]{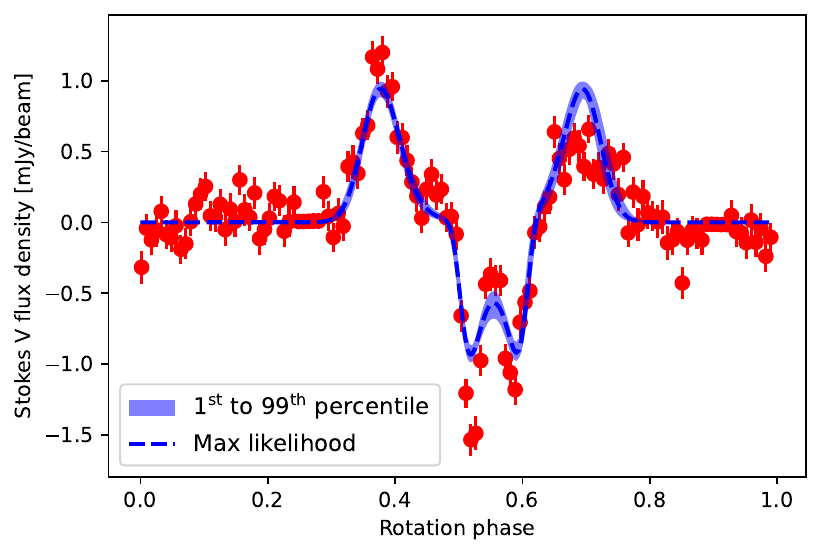}
    \caption{Fit to the MeerKAT burst a (see Figures \ref{fig:MKT_DS_V} and \ref{fig:MKT_DS_zooms}) using the framework described in \protect \cite{kavanagh2024unravelling}. The best-fitting light curve is sampled from the posterior distributions of the model parameters (see Figure \ref{fig:modelling_corner}).}
    \label{fig:modelling_lc}
\end{figure}
In this Section we follow the method used to model the magnetic field of WISE J$062309.94-045624.6$ by \cite{kavanagh2024unravelling}, and apply their \textsc{charm}\footnote{\url{https://github.com/robkavanagh/charm}} software to \bdshort. \cite{kavanagh2024unravelling} present a framework where a single field line in a UCD dipolar magnetic field is chosen to emit ECME. The dipolar magnetic field is characterised by the magnetic obliquity $\beta$, the angle between the magnetic axis and the star rotation axis and the inclination $i$, the angle between the rotation axis and the observer line-of-sight. Each active field line can be ascribed a latitude ($\theta$) and longitude ($\phi$), defined with respect to a rotation phase offset ($\phi_0$). Finally, each active field line will produce one ECME emission cone, with an opening angle $\alpha$ and a thickness of $\Delta \alpha$. We refer the reader to \cite{kavanagh2024unravelling} for a detailed sketch of this setup. 

We apply the \textsc{charm} active field line framework to \bdshort. We choose the UHF light curve of burst a in Figure \ref{fig:MKT_DS_zooms}, as it has a representative shape and is not interrupted by a calibrator scan. We note that this differs from the approach in \cite{kavanagh2024unravelling}, who perform this analysis on a phase-folded average light curve. We tried this approach, but found that it was highly sensitive to specific choices in the phase-folding or cross-correlation setup, because the bursts from \bdshort\ have a much shorter duty cycle compared to WISE J$062309.94-045624.6$ in \cite{kavanagh2024unravelling}. We therefore opt to perform this analysis on a single burst instead, so the fitting algorithm can pick up on the sharp features present in the single burst profile. 
To get a sense of the magnetic field setup, we choose a single active field line. We note that a simple single-field-line model cannot account for the polarization flip observed in burst b (Figure~\ref{fig:MKT_DS_V}). If this reversal arises from emission originating at the opposite footpoint of the same magnetic field line, it would be expected to occur at a different rotational phase. A more realistic description of the magnetic topology of \bdshort\ may require multiple active field lines or auroral rings \citep{bloot2024phenomenology}. Such more complex magnetic geometries could potentially reproduce the polarization reversal seen in burst b in Figure \ref{fig:MKT_DS_V} \citep[see, e.g.,][]{pineda2017panchromatic}. However, a detailed modelling effort is beyond the scope of this work and would require a longer (MeerKAT) monitoring campaign. Table \ref{tab:modelling_table} shows the parameters we obtain, and Figure \ref{fig:modelling_lc} shows the light curve in red, with a fit sampled from the posteriors overlaid in blue. We show the full corner plot in Figure \ref{fig:modelling_corner} in \ref{app:corner_plot}.

\begin{table}
\centering
\begin{tabular}{l c}
\hline
Parameter & Value \\
\hline
$\cos i$ & $0.67^{+0.11}_{-0.16}$ \\
$i$ [$^{\circ}$] & $47.9^{+11.4}_{-9.2}$ \\
$\beta$ [$^{\circ}$] & $40.1^{+9.2}_{-10.9}$ \\
$\phi_0$ [$^{\circ}$] & $338.6^{+0.6}_{-0.6}$ \\
$\theta$ [$^{\circ}$] & $32.7^{+5.2}_{-7.4}$ \\
$\phi$ [$^{\circ}$] & $181.6^{+0.6}_{-0.6}$ \\
$F_{0}$ [mJy] & $0.95^{+0.03}_{-0.03}$ \\
$\alpha$ [$^{\circ}$] & $43.5^{+11.5}_{-8.6}$ \\
$\Delta \alpha$ [$^{\circ}$] & $1.72$ \\
\hline
\end{tabular}
\caption{Model parameters with asymmetric uncertainties. Angular quantities are reported in degrees.\label{tab:modelling_table}}
\end{table}

To build some intuition for this configuration, consider the range of angles the magnetic axis sweeps out relative to the line of sight over one rotation. For a dipole tilted by an obliquity $\beta$ from the rotation axis, itself inclined by $i$ to the line of sight, this angle varies between $|i-\beta|$ and $i+\beta$. With $i \approx 48^{\circ}$ and $\beta \approx 40^{\circ}$, the range is approximately $8^{\circ}$--$88^{\circ}$; at one rotation phase the magnetic pole points almost directly at the observer, while half a rotation later it lies almost in the plane of the sky. The modelled field line, offset from the pole by $\theta \approx 33^{\circ}$, sweeps a similarly wide range of viewing angles. Since the emission cone is both narrow ($\alpha \approx 44^{\circ}$) and thin ($\Delta\alpha \approx 1.7^{\circ}$), we observe relatively short, sharp bursts as seen in Figure \ref{fig:MKT_DS_V} rather than a broad, slowly-varying profile.

The cone parameters carry information about the emitting electrons. ECME relies on a horseshoe- or loss-cone-shaped velocity distribution, formed as low-pitch-angle electrons precipitate while high-pitch-angle electrons are mirrored back along the field line. In the frame co-moving with the electrons' parallel motion, maser growth peaks near $90^{\circ}$ from the local field, giving a hollow-cone beam; relativistic aberration then 'tips this forward' by an amount set by $v_{\parallel}/c$ \citep[e.g.][]{melrose1982electron,treumann2006electron}. A cone offset well away from $90^{\circ}$, as found here ($\alpha \approx 44^{\circ}$), points to electrons with a non-negligible field-aligned velocity ($v_{\parallel}/c \approx \cos\alpha \approx 0.7$)\footnote{A field aligned velocity $v_{\parallel}/c\approx 0.7$ implies $\gamma\approx1.4$ and thus a minimum kinetic energy of $(\gamma -1)\times 511 \; \rm{keV} \approx 205 \; \rm{keV}$ per electron. This estimate is consistent with expected electron energies for ECME \citep[between 10 keV and 1 MeV;][]{dulk1985radio}. Note that this is a lower bound since we only consider one velocity component.}. Such a substantial field-aligned velocity component can be supplied by an electric field that develops along the magnetic field line itself and accelerates electrons downward; the same process is thought to energise auroral electrons around Earth and Jupiter \citep{ergun2000electron,mauk2017discrete}. The small thickness, $\Delta\alpha \approx 1.7^{\circ}$, further implies a narrow spread in resonant electron energies and pitch angles.

As demonstrated by \citet{kavanagh2024unravelling}, the duty cycle of auroral radio bursts is strongly dependent on viewing geometry. UCDs observed equator-on with a small magnetic obliquity produce the narrowest bursts, whereas those viewed closer to pole-on with large magnetic obliquities exhibit much broader burst profiles. \bdshort\ is an example of the former configuration, while WISE J$062309.94-045624.6$ exemplifies the latter \citep{kavanagh2024unravelling}. Consequently, radio surveys are biased against detecting sources with geometries similar to \bdshort, particularly when searches rely on time-averaged radio flux integrated over one or more rotation periods. This effect is evident in Figure \ref{fig:askap_lightcurve}: averaging over the 2-hour 855 MHz observation dilutes the burst emission and results in only a marginal detection of \bdshort, whereas the 12-minute 887 MHz observations centred on the burst yield bright, high-significance detections.

\begin{table*}[h]
    \hspace*{-0.4cm}
    \begin{tabular}{c|ccc|c|cc|c} \hline
 &  &  &  &  &  $log(L_{peak})$ & Circular & Radio  \\
Name & SpT & log($L_{{bol}}/L_{\odot})$ & T [K] & d [pc] & [erg/s/Hz] & polarisation \% & period [hr] 
\\\hline
2MASS J$13142039+1320011$ & M7$^{a}$ & $-2.631 \pm 0.01$ & $2947 \pm 4$ $^{[1]}$ & $17.249 \pm 0.013$ $^{[1]}$ & $15.4$ & $100$ $^{[2,3,4]}$ &$3.89 \pm 0.05$ $^{[3]}$\\
TVLM$513-46546$ & M8.5 & $-3.57 \pm 0.01$ & $2242 \pm 55$ $^{[5]}$ & $10.733 \pm 0.016$ $^{[6]}$ & $14.8$ & $100$ $^{[2,7,8,9]}$ &$1.96$ $^{[7]}$\\
LSR J$1835+3259$ & M8.5 & $-3.5 \pm 0.0$ & $2316 \pm 51$ $^{[5]}$ & $5.689 \pm 0.002$ $^{[10]}$ & $14.0$ & $100$ $^{[2,11,12]}$ &$2.84 \pm 0.01$ $^{[12]}$\\
2MASS J$07464256+2000321$ & L1.5$^{b}$ & $-3.77 \pm 0.02$ & $2000 \pm 200$ $^{[13]}$ & $12.219 \pm 0.045$ $^{[14]}$ & $15.3$ & $100$ $^{[2,15]}$ &$2.076 \pm 0.002$ $^{[15]}$\\
2MASS J$0036159+182110$ & L3.5 & $-3.93 \pm 0.01$ & $1869 \pm 64$ $^{[5]}$ & $8.736 \pm 0.011$ $^{[10]}$ & $13.8$ & $>60$ $^{[2,12,8]}$ &$3.08 \pm 0.05$ $^{[12]}$\\
SIMP J$013656.5+093347$ & T2.5 & $-4.688 \pm 0.005$ & $1098 \pm 6$ $^{[16]}$ & $6.139 \pm 0.037$ $^{[17]}$ & $13.35$ & $64 \pm 16.0$ $^{[18]}$ &$2.88 \pm 0.034^{c}$ $^{[19]}$\\
2MASS J$10430758+2225236$ & L8 & $-4.53 \pm 0.1$ & $1301 \pm 109$ $^{[19,21]}$ & $16.4 \pm 3.2$ $^{[19]}$ & $13.2$ & $61 \pm 11.7$ $^{[19]}$ &$2.21 \pm 0.14$ $^{[19]}$\\
WISEPC J$112254.72+255022.2$ & T6 & $-5.38 \pm 0.05$ & $790 \pm 58$ $^{[21]}$ & $15.1 \pm 0.5$ $^{[22]}$ & $14.5$ & $100$ $^{[23]}$ &$1.95 \pm 0.03$ $^{[23]}$\\
2MASS J$22282889-4310262$ & T6 & $-5.28 \pm 0.07$ & $891 \pm 82$ $^{[5]}$ & $10.9 \pm 0.3$ $^{[24]}$ & $13.67$ & $>50$ $^{[25]}$ &$1.43 \pm 0.16^{d}$ $^{[26]}$\\
2MASS J$12373919+6526148$ & T6.5 & $-5.36 \pm 0.04$ & $851 \pm 74$ $^{[5]}$ & $10.42 \pm 0.52$ $^{[19]}$ & $13.05$ & $65 \pm 15.0$ $^{[19]}$ &$2.28 \pm 0.1$ $^{[19]}$\\
2MASSI J$1047539+212423$ & T6.5 & $-5.3 \pm 0.04$ & $880 \pm 76$ $^{[5]}$ & $10.6 \pm 0.4$ $^{[27]}$ & $13.2$ & $80 \pm 20.0$ $^{[2,28,18]}$ &$1.77 \pm 0.04$ $^{[28]}$\\
WISE J$062309.94-045624.6$ & T8 & $-5.77 \pm 0.11$ & $624 \pm 40$ $^{[21]}$ & $11.44 \pm 0.37$ $^{[22]}$ & $14.8$ & $66 \pm 9.0$ $^{[29]}$ &$1.912 \pm 0.005$ $^{[29]}$\\ \hline
WISE J$151721.13+052929.3$ & T8 & $-5.77 $ & $755 \pm 150$ $^{[30]}$ & $21.7 \pm 1.0$ $^{[31]}$ & $15.75$ & $20 \pm 3.0$ $^{[32]}$ &$1.149 \pm 0.002$ $^{[32]}$ \\ \hline
    \end{tabular}
    \caption{Properties of UCDs (spectral type M7 or later) showing periodic radio emission. This paper describes the radio emission from \bdfull\ (last row) in detail. A machine-readable version of this Table is included in the reproduction package; see the data availability statement. References for the spectral types and bolometric luminosities are the same as the temperature reference. References for the peak radio luminosities are the same as the reference for the circular polarisation fraction. \\
    $^{a}$: The radio emission originates from the secondary in this close-in M7+M7 binary \citep{forbrich2016high}. \cite{williams2015simultaneous} identify two different but similar periodicities in the radio emission of this source, possibly linked to two different radio-emitting spots on one object with differential rotation, or both components of the binary rotate with very similar periodicities. \\
    $^{b}$: The radio emission originates from the secondary in this close-in L0+L1.5 binary \citep{harding2013spin}. \\
    $^{c}$: The radio period in \cite{kao2018strongest} is tentative because there is only one clear burst. We still adopt it here, as it is in agreement with observations from \cite{kao2016auroral} and the photometric period \citep{artigau2009photometric}. \\
    $^{d}$: Listed here is the mid-infrared photometric period. Tentative radio periodicity (occurrence intervals of 47 min and 58 mins \protect \citep{wandia2026radio}) that could correspond to this photometric period. We assume a similar bolometric luminosity as for WISE J062309.94-045624.6. \\ References: [1]:\cite{dupuy2016high}, [2]:\cite{pineda2017panchromatic}, [3]:\cite{mclean2011periodic}, [4]:\cite{williams2015simultaneous}, [5]:\cite{filippazzo2015fundamental}, [6]:\cite{gawronski2017physical}, [7]:\cite{hallinan2007periodic}, [8]:\cite{mclean2012radio}, [9]:\cite{wolszczan2014timing}, [10]:\cite{dahn2017ccd}, [11]:\cite{berger2008simultaneous}, [12]:\cite{hallinan2008confirmation}, [13]:\cite{harding2013spin}, [14]:\cite{dieterich2014solar}, [15]:\cite{berger2009periodic}, [16]:\cite{gagne2017simp}, [17]:\cite{weinberger2016trigonometric}, [18]:\cite{kao2016auroral}, [19]:\cite{kao2018strongest}, [20]:\cite{kao2018strongest,sanghi2023hawaii}, [21]:\cite{sanghi2023hawaii}, [22]:\cite{kirkpatrick2019preliminary}, [23]:\cite{guirado2025main}, [24]:\cite{smart2013nparsec}, [25]:\cite{wandia2026radio}, [26]:\cite{clarke2008search}, [27]:\cite{vrba2004preliminary}, [28]:\cite{williams2015rotation}, [29]:\cite{rose2023periodic}, [30]:\cite{mace2013study}, [31]:\cite{kirkpatrick2021field}, [32]:this work}
    \label{tab:periodic_UCDs}
\end{table*}

\section{Periodic radio emission from UCDs} \label{sec:UCD_context}
\subsection{A sample of UCDs with periodic radio emission}
To place \bdshort\ in the context of the broader UCD population, we compile here the complete sample of UCDs with confirmed periodic radio emission (Table \ref{tab:periodic_UCDs}), and review some of the key properties.

As noted in the introduction, the discovery of periodic, highly polarised radio bursts from TVLM 513$-$46546 demonstrated that ECME is responsible for the coherent radio emission observed in ultracool dwarfs \citep{hallinan2007periodic,hallinan2008confirmation}. In addition to its bursting component, TVLM 513$-$46546 exhibits a non-varying, weakly polarised component, interpreted either as depolarised coherent emission or as gyrosynchrotron radiation from a more extended magnetospheric region \citep{hallinan2006rotational}. Long-baseline astrometry rules out companions more massive than Jupiter on orbits longer than one year, strengthening the interpretation of an intrinsic auroral process \citep{gawronski2017physical}, though \cite{curiel2020astrometric} find evidence for a planetary companion candidate.

Periodic radio emission has since been confirmed in several other UCDs spanning late-M, L and T spectral types. We include a source in Table \ref{tab:periodic_UCDs} if it shows radio bursts that are periodic and more than $50\%$ circularly polarised, for which ECME is the most likely emission mechanism. We note that for some systems (2MASS J13142039$+$1320011; \citep{mclean2011periodic, williams2015simultaneous} and TVLM 513$-$46546; \citep{hallinan2007periodic, hallinan2006rotational}) bright polarised bursts are sometimes observed and sometimes excluded at this significance level. For the binary systems NLTT 33370 and 2MASS J$07464256+2000321$, the radio emission has been pinpointed to one of the binary constituents using VLBI \citep{forbrich2016high, harding2013spin}. We highlight the properties of individual sources below, and detail our reasons for excluding some candidate systems.

The L3.5 dwarf 2MASS J00361617$+$1821104 displays coherent, highly circularly polarised radio bursts with a rotation period of $3.08\pm0.05$ hr \citep{hallinan2008confirmation}, with polarisation fractions up to 60\% reported in earlier observations \citep{berger2002flaring,berger2005magnetic}. The source also exhibits a weaker interpulse whose low apparent polarisation fraction was initially attributed to depolarisation or mode conversion of the maser emission in a lower-density cavity \citep{hallinan2006rotational}. A recent multi-epoch VLA and VLBI study by \citet{martin-carrero2026} reveals a more complex polarisation morphology: the circular polarisation light curves shows peaks of both left- and right-hand polarisation, indicative of ECME emission from opposite hemispheres with different circular polarisation handedness. \citeauthor{martin-carrero2026} model the radio emission with a hybrid auroral-ring plus active-field-line scenario using the \textsc{charm} geometric framework, similar to Section~\ref{sec:discussion_afl} of this work. Similarly, the M8.5 dwarf LSR J1835$+$3259 shows coherent radio bursts with a period of $2.84\pm0.01$ hr, accompanied by a quiescent unpolarised component \citep{hallinan2008confirmation}. Both 2MASS J00361617$+$1821104 and LSR J1835$+$3259 were initially detected in radio surveys targeting active late-M dwarfs \citep{berger2002flaring}. Additional periodic sources include the L-dwarf binary 2MASS J07464256$+$2000321, which displays 100\% circular polarisation and a rotation period of $124.32\pm0.11$ min \citep{antonova2008mini,berger2009periodic,lynch2015wideband}.

The first radio-emitting T dwarf was detected in a large survey of 27 UCDs with the Arecibo radio telescope \citep{route2016second,route2016radio}. This T6.5 dwarf, 2MASSI J1047539$+$212423, exhibits periodic radio emission with a period of $1.77\pm0.04$ hr \citep{williams2015rotation}. Two sources in Table \ref{tab:periodic_UCDs} have a more tentative radio period. SIMP J$013656.5+093347$ has only been detected to burst twice, in a single observation \citep{kao2016auroral}; a separate study \citep{kao2018strongest} finds the radio period listed in Table \ref{tab:periodic_UCDs} despite these data showing only one significant burst, which we adopt here as it agrees with both the \cite{kao2016auroral} observations and the photometric period \citep{artigau2009photometric}. The T6 dwarf 2MASS J22282889$-$4310262 shows highly circularly polarised ($50$--$80$\%) radio bursts with occurrence intervals of 47 and 58 minutes over two observing epochs \citep{wandia2026radio}; these observations cannot conclusively determine a radio period, but are consistent with (harmonics of) the infrared photometric period, which we list in Table \ref{tab:periodic_UCDs}. The T8 dwarf WISE J062309.94$-$045624.6 similarly shows rotationally modulated, circularly polarised radio emission, with a period of $1.912\pm0.005$ hr \citep{rose2023periodic}.

We exclude the T-dwarf binary WISEP J101905.63$+$652954.2 from Table \ref{tab:periodic_UCDs}, despite reports of auroral radio emission \citep{vedantham2023polarised}, because this emission cannot yet be uniquely attributed to one of the binary components: the detection of two distinct periodicities (3 hr and 0.79 hr) suggests both brown dwarfs may contribute, but unresolved observations and the absence of independent rotation constraints prevent a definitive assignment \citep{yiu2025unusual}.

Finally, we exclude two well-known radio-bursting UCDs for which no radio period has been established: the M9 dwarf LP $944-20$ \citep{berger2001discovery}, as mentioned in the introduction, and the M9.5 dwarf BRI $0021-0214$ \citep{berger2002flaring}. Both systems show only weak or marginal circular polarisation ($\sim30\%$), underscoring the diversity of radio phenomenology among UCDs and the value of the polarisation and periodicity criteria used to build our comparison sample.

\subsection{\bdshort\ in context: low circular polarisation and high luminosity}

\begin{figure}
    \centering
    \includegraphics[width=\linewidth]{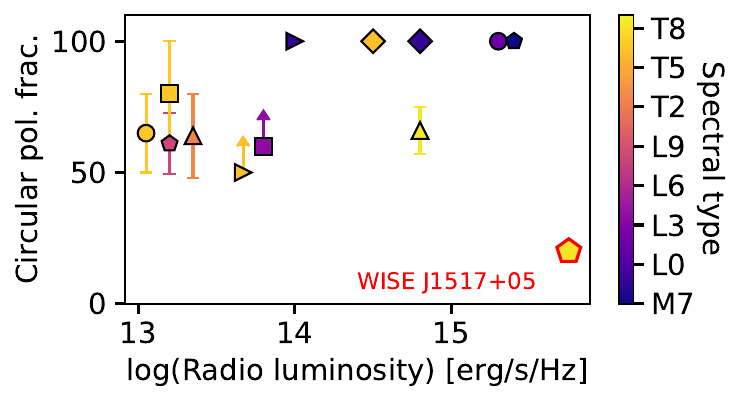}
    \caption{Circular polarisation fraction as a function of the spectral radio luminosity of the bursting emission for the periodic UCDs in Table \ref{tab:periodic_UCDs}. \bdshort\ stands out because its bursts have low circular polarisation and a high luminosity.}
    \label{fig:pol_lum}
\end{figure}

The radio properties of \bdshort\ distinguish it from the currently known sample of periodically emitting UCDs: it is the most radio-luminous periodic UCD detected to date, and its measured radio period places it among the fastest-rotating UCDs known. Furthermore, \bdshort\ exhibits an unusually low degree of circular polarisation compared to the rest of the periodic UCD population (Figure~\ref{fig:pol_lum}), providing a rare opportunity to investigate the origin of polarisation diversity in auroral radio emitters.

\subsection{\bdshort\ in context: rapidly rotating UCDs}

\begin{figure*}
    \centering
    \includegraphics[width=\linewidth]{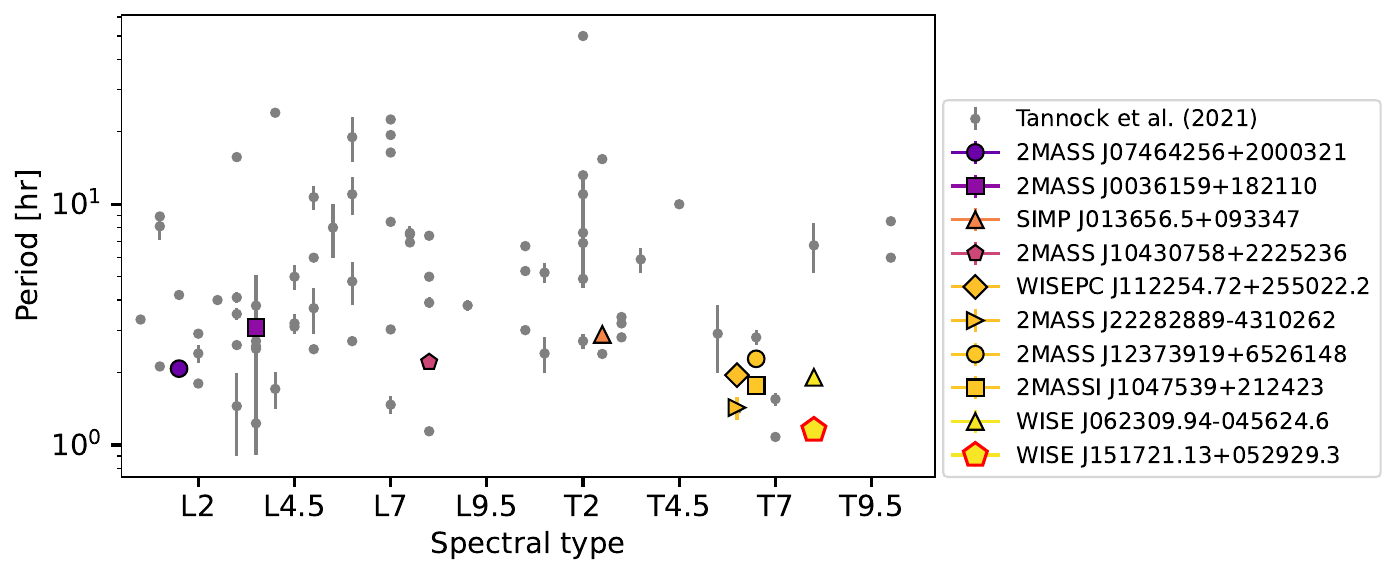}
    \caption{Rotation period as a function of spectral type for a sample of 78 L0–Y0 dwarfs with confirmed photometric rotation periods from \protect\cite{tannock2021weather} in grey. Overplotted are the L- and T-type radio-bursting periodic UCDs, as summarised in Table \ref{tab:periodic_UCDs}. }
    \label{fig:UCD_period_context}
\end{figure*}

Based on a small sample of five radio-bursting UCDs, \cite{kao2018strongest} pose the idea that rapid rotation is important in producing strong dipolar magnetic fields in UCDs and/or rapid rotation is a key ingredient for driving the currents necessary for ECME. \bdshort\ is a valuable addition to this sample of UCDs with confirmed periods, as presented in Table \ref{tab:periodic_UCDs}. We note here that we assume the radio period to be the rotation period, as the modulation of the radio emission is driven by the rotation of the star in an ECME scenario. In Figure \ref{fig:UCD_period_context} we show the rotation period as a function of spectral type for L, T, and Y-dwarfs. The grey dots show a sample of 78 L0–Y0 dwarfs with confirmed photometric rotation periods from \cite{tannock2021weather}. The coloured points show the radio-bursting L and T dwarfs as presented in Table \ref{tab:periodic_UCDs}. This figure supports the hypothesis that rapid rotation is a key ingredient for ECME, as all radio-detected UCDs are in the lower period range compared to the full sample of UCDs.\\

We also examine possible correlations between radio period, spectral type, and radio burst luminosity for the objects listed in Table \ref{tab:periodic_UCDs}. Different theoretical frameworks predict distinct scalings between radio power, rotation rate and magnetic field strength \citep{zarka2007plasma, nichols2012origin, turnpenney2017auroral}. In our extended sample, we find no significant correlation between radio burst spectral luminosity and spectral type or radio period, see \ref{app:correlations}. This is consistent with earlier work by \cite{mclean2012radio} and \cite{pineda2017panchromatic}, where it was suggested that rapid rotation is a threshold condition to power ECME. Once this regime is reached, certain saturation effects may come into play. The conductance limit of the UCDs imposes a saturation on the 
total current that can be sustained, which imposes an upper limit on the power of the radio emission. Additionally, an increased power of ECME hinges on a readily available plasma supply, which may be a limiting factor in the increase of the radio burst luminosity. 

Finally, we emphasise that the magnetic field strength (which factors into the radio luminosity) is not expected to increase with rotation period. UCDs operate in the saturated dynamo regime \citep{pizzolato2003stellar, wright2010wide, mclean2012radio}: their extremely short rotation periods and long convective turnover times imply Rossby numbers of order $10^{-3}$–$10^{-2}$, well below the canonical saturation threshold of $\sim 0.1$ \citep{xiang2026magnetic}. In this regime, magnetic field generation no longer increases with rotation rate, removing the secondary driver for any luminosity–period correlation. We note that \citet{kao2018strongest} measured kG-level ($3.2$--$6.2$\,kG) fields on rapidly-rotating brown dwarfs from their ECME spectral cutoffs and found that age, mass, and effective temperature could not account for the field strengths produced. This implies that saturated-regime dynamo scaling relations underpredict the observed magnetic field strengths, indicating a gap in our theoretical understanding of these objects. Our results reinforce the interpretation that ECME emission in UCDs is governed by threshold and saturation effects, rather than a continuous scaling with rotation or magnetic field strength, although the theoretical framework for such scalings in UCDs remains incomplete.

\section{Conclusion} \label{sec:conclusion}

We report the discovery of radio bursts from the T8 brown dwarf \bdfull\ produced by auroral ECME. \bdshort\ is a valuable addition to the small sample of radio-bursting UCDs (Table \ref{tab:periodic_UCDs}). Notably, \bdshort\ has the fastest rotation period of any radio-detected UCD to date, strengthening the hypothesis that rapid rotation plays a key role in generating strong dipolar magnetic fields in convective dynamos \citep{kao2018strongest}. The radio emission from \bdshort\ is most likely produced by ECME, although the low circular polarisation fractions are atypical, but can be explained by propagation effects. We show that the radio emission from \bdshort\ can be well modelled with a single active field line using \textsc{charm}, a framework for characterising magnetic fields via Bayesian inference from radio-emitting objects such as stars, brown dwarfs, and planets \citep{kavanagh2024unravelling}. 

Currently, the polarisation flip observed in one of the \bdshort\ bursts (see Figure~\ref{fig:MKT_DS_V}) remains unexplained. Bursts of opposite handedness are typically attributed to emission from opposite regions of the magnetosphere; however, in this case the burst occurs at the same rotational phase as the main burst, making an interpulse interpretation unlikely \citep{hallinan2008confirmation}.

\bdshort\ is the most radio-luminous UCD detected to date, indicating that the distance out to which we can detect radio-bursting UCDs may be larger than previously assumed. The aforementioned relation between radio activity and rapid rotation reduces the likelihood that bursts are missed due to limited observing durations. Together, the high spectral luminosity and preference for short rotation periods suggest that more UCDs may be discovered in untargeted radio surveys than previously anticipated \citep[see e.g. Section~4.3 of][]{rose2023periodic}.  We note that a search for highly circularly polarised sources (filtering on the ratio of circular polarisation and total intensity), may miss UCDs with low circular polarisation fractions, similar to \bdshort.

\section*{Data availability}

The data products presented in this paper are available in a reproduction package via Zenodo, at \url{https://dx.doi.org/10.5281/zenodo.21869650}. This reproduction package contains the dynamic spectrum files and all input files necessary to create the figures in this paper. Jupyter notebooks and conda environments are also provided.

The raw data products for the ASKAP observations are available through CSIRO’s ASKAP Science Data Archive \citep[CASDA;][]{huynh2020csiro}\footnote{\url{casda.csiro.au}}. The project codes of the various ASKAP observations are available in the reproduction package. The ATCA observations can be obtained under project code C3363 on the Australia Telescope Online Archive\footnote{\url{http://atoa.atnf.csiro.au}}. The MeerKAT observations are available on the MeerKAT archive\footnote{\url{https://archive.sarao.ac.za/observations}} under project code DDT-20251022-ID-01.

\section*{Acknowledgements}
We thank Manisha Caleb for her help in preparing the MeerKAT observational setup, Rob Kavanagh for making his \textsc{charm} modelling code open-access and user-friendly, and Qichen Huang for valuable discussions. We thank the MeerKAT scheduling team for promptly observing \bdshort, and for the opportunity to repeat our observations because the first observations were affected by an X-class solar flare. We thank the anonymous reviewer for their careful reading of the manuscript and constructive comments, which have helped improve the quality and clarity of this work.

This research was supported by the Australian Research Council Centre of Excellence for Gravitational Wave Discovery (OzGrav), project number CE230100016. DK was supported by NSF grant AST-2511757. 

The MeerKAT telescope is operated by the South African Radio Astronomy Observatory, which is a facility of the National Research Foundation, an agency of the Department of Science and Innovation.

This scientific work uses data obtained from Inyarrimanha Ilgari Bundara, the CSIRO Murchison Radio-astronomy Observatory. We acknowledge the Wajarri Yamaji People as the Traditional Owners and Native Title Holders of the Observatory site. CSIRO’s ASKAP radio telescope is part of the Australia Telescope National Facility (\url{https://ror.org/05qajvd42}). Operation of ASKAP is funded by the Australian Government with support from the National Collaborative Research Infrastructure Strategy. ASKAP uses the resources of the Pawsey Supercomputing Research Centre. Establishment of ASKAP, Inyarrimanha Ilgari Bundara, the CSIRO Murchison Radio-astronomy Observatory and the Pawsey Supercomputing Research Centre are initiatives of the Australian Government, with support from the Government of Western Australia and the Science and Industry Endowment Fund.

The Australia Telescope Compact Array is part of the Australia Telescope National Facility (\url{https://ror.org/05qajvd42}) which is funded by the Australian Government for operation as a National Facility managed by CSIRO. We acknowledge the Gomeroi people as the Traditional Owners of the Observatory site.

 This publication makes use of data products from the Wide-field Infrared Survey Explorer (WISE), which is a joint project of the University of California, Los Angeles, and the Jet Propulsion Laboratory/California Institute of Technology, funded by the National Aeronautics and Space Administration. 
Microsoft Copilot and Claude (Sonnet 4.6) were used for editorial assistance, specifically improving clarity and phrasing in parts of the manuscript. All scientific content, analysis, and conclusions are the authors’ own.


\bibliography{bibtemplate}

\appendix
\renewcommand{\thefigure}{\Alph{section}\arabic{figure}}
\setcounter{figure}{0}
   
\section{Corner plot \textsc{charm}} \label{app:corner_plot}
\begin{figure*}
    \centering
    \includegraphics[width=\linewidth]{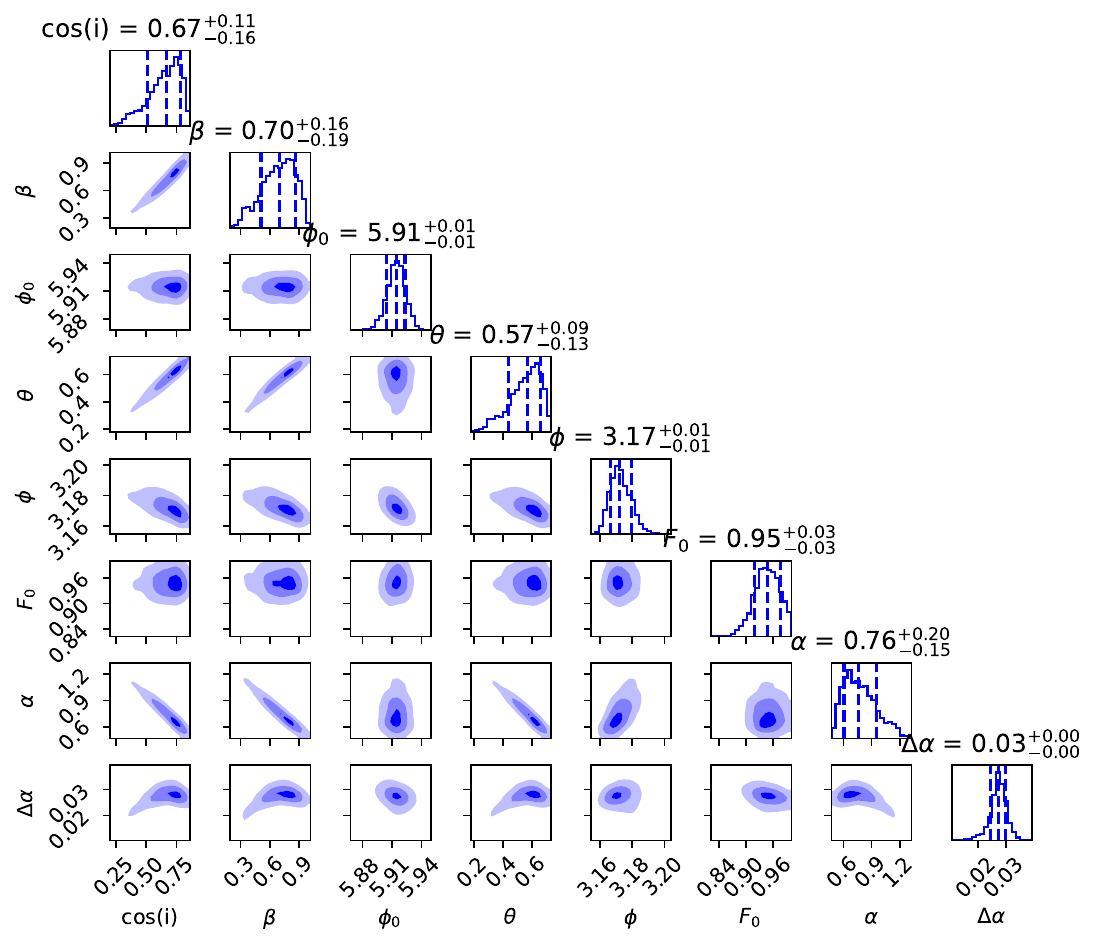}
    \caption{Posterior distributions of the best-fitting model parameters listed in Table \ref{tab:modelling_table} plotted using \textsc{corner} \citep{foreman2016corner}. The shaded regions in each panel, bar the diagonal ones, show the 16th, 50th, and 84th percentiles going from darkest to lightest in colour. The contours are smoothed by setting the smooth parameter to 1 in the corner function. The diagonal panels show the one-dimensional histograms for each parameter (unsmoothed).}
    \label{fig:modelling_corner}
\end{figure*}

Figure \ref{fig:modelling_corner} shows the posterior distributions of the best-fitting model parameters listed in Table \ref{tab:modelling_table}. Note that Figure \ref{fig:modelling_corner} shows the angles in radians, whereas Table \ref{tab:modelling_table} shows the angles in degrees. The model includes a single active field line and was applied to the MeerKAT UHF light curve of burst a (Figure \ref{fig:MKT_DS_zooms}), as described in detail in Section \ref{sec:discussion_afl}.

\setcounter{figure}{0}
\section{Exploring the periodic radio UCD sample} \label{app:correlations}
\begin{figure*}
    \centering
    \includegraphics[width=\linewidth]{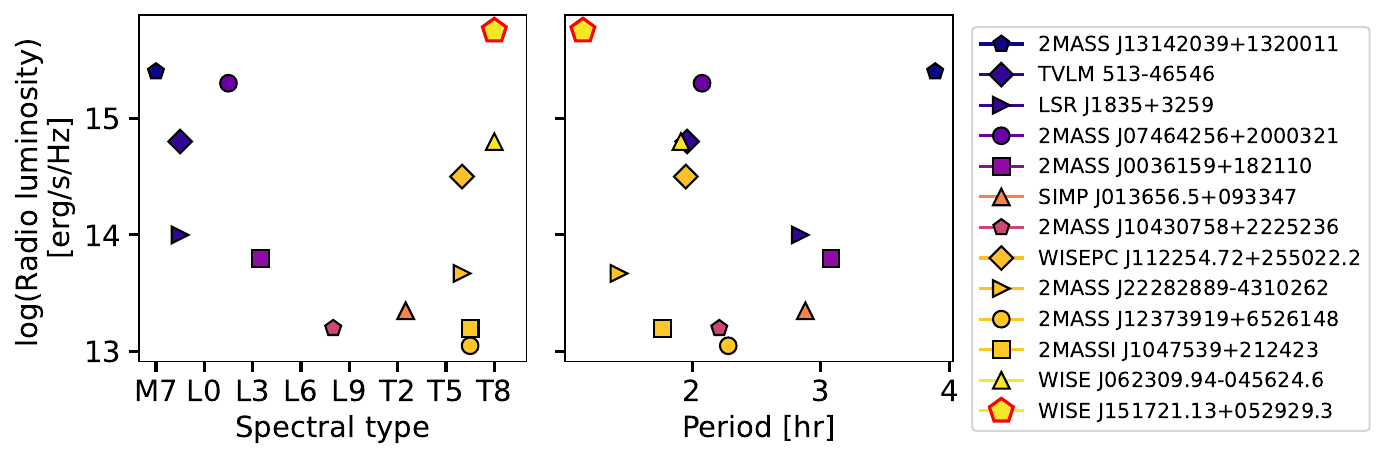}
    \caption{Radio luminosity as a function of spectral type (left) and as a function of radio period (right) of the bursting emission for the periodic radio UCD sample in Table \ref{tab:periodic_UCDs}. }
    \label{fig:correlations_UCS_sample}
\end{figure*}

Figure \ref{fig:correlations_UCS_sample} shows the radio luminosity as a function of spectral type (left) and as a function of radio period (right) of the bursting emission for the periodic radio UCD sample in Table \ref{tab:periodic_UCDs}. There is no statistically significant correlation between either of these quantities; the Pearson correlation coefficients are $\mathrm{P}=-0.27$ and $\mathrm{P}=-0.056$ with p-values of $0.37$ and $0.85$ for the left and right plot, respectively.

\end{document}